\documentclass{article}

\usepackage{spconf,amsmath,epsfig}
\usepackage{cite, graphicx, psfrag, subfigure, url, stfloats, amsmath, amssymb}

\DeclareMathAlphabet{\mathpzc}{OT1}{pzc}{m}{it}

\newtheorem{lemma}{Lemma}

\title{Spatial-Spectral Joint Detection for Wideband Spectrum Sensing in Cognitive Radio Networks}
\name{Zhi Quan$^{\dag}$, Shuguang Cui$^{\ddag}$, Ali H.
Sayed$^{\dag}$, and H. Vincent Poor$^{\S}$
\thanks{This research was supported in part by NSF under Grants ANI-03-38807, CNS-06-25637, ECS-06-01266,
ECS-07-25441, CNS-06-25637, and by DoD under Grant
HOTRN-07-1-0037.} }
\address{\normalsize $^{\dag}$Department of Electrical Engineering, University of California, Los Angeles,
CA 90095 \\
\normalsize$^{\ddag}$Department of Electrical and Computer Engineering, Texas A\&M University, College Station, TX 77843 \\
\normalsize$^{\S}$Department of Electrical Engineering, Princeton University, Princeton, NJ 08544 \\
\normalsize Email: \{quan, sayed\}@ee.ucla.edu, cui@tamu.edu,
poor@princeton.edu }
\begin{document}
%
\maketitle
\begin{abstract}

Spectrum sensing is an essential functionality that enables
cognitive radios to detect spectral holes and opportunistically
use under-utilized frequency bands without causing harmful
interference to primary networks. Since individual cognitive
radios might not be able to reliably detect weak primary signals
due to channel fading/shadowing, this paper proposes a cooperative
wideband spectrum sensing scheme, referred to as
\emph{spatial-spectral joint detection}, which is based on a
linear combination of the local statistics from spatially
distributed multiple cognitive radios. The cooperative sensing
problem is formulated into an optimization problem, for which
suboptimal but efficient solutions can be obtained through
mathematical transformation under practical conditions.


\end{abstract}
\vspace{-6pt}
\section{Introduction}
\label{sec:intro}

\vspace{-8pt}

As an essential functionality of cognitive radio (CR)
 networks \cite{MitolaIII95}, spectrum sensing needs to reliably
detect weak primary radio signals of possibly-unknown formats.
Generally, spectrum sensing techniques can be classified into
three categories: energy detection, matched filter coherent
detection \cite{Poor1994}, and cyclostationary feature detection.
Since non-coherent energy detection is simple and able to generate
the spectrum-occupancy information quickly, we adopt it as the
building block for constructing the proposed wideband spectrum
sensing schemes.

The literature on wideband spectrum sensing for CR networks is
limited. An earlier approach is to use a tunable narrowband
bandpass filter at the RF front-end to sense one narrow frequency
band at a time, over which the existing narrow-band spectrum
sensing techniques can be applied. In order to search over
multiple frequency bands at a time, the RF front-end needs a
wideband architecture, and spectrum sensing usually operates over
an estimate of the power spectral density (PSD) of the wideband
signal. In \cite{Tian2006}, wavelet transformation was used to
estimate the PSD over a wide frequency range given its
multi-resolution features. However, no prior work attempts to make
decisions over multiple frequency bands jointly, which is
essential for implementing efficient CR networks.

In this paper, we consider the situation in which spectrum sensing
is compromised by destructive channel conditions between the
target-under-detection and the detecting cognitive radios, which
makes it hard to distinguish between a white spectrum and a weak
signal. We propose a cooperative wideband spectrum sensing scheme
that exploits the spatial diversity among cognitive radios to
improve the sensing reliability. The cooperation is based on a
linear combination of local statistics from spatially distributed
cognitive radios \cite{Quan2007c}\cite{Quan2007J}, where the
signals are assigned different weights according to their positive
contributions to joint sensing. In such a scenario, we treat the
design of distributed wideband spectrum sensing as a
\emph{spatial-spectral joint detection} problem, which is further
formulated into an optimization problem with the objective of
maximizing the overall opportunistic throughput under constraints
on the interference to primary users. Through mathematical
reformulation, we derive a suboptimal but efficient solution for
the optimization problem, which can considerably improve sensing
performance.

\vspace{-10pt}

\section{System Model}\label{sec:model}
\label{sec:model}

\vspace{-8pt}

\begin{figure}
\centering
\includegraphics[width=2.6in]{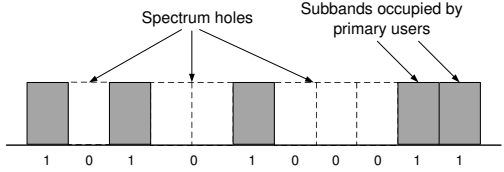} \vspace{-10pt}
\small  \caption{\small Illustration of the occupancy of a
multiband channel.} \label{fig:sppool} \vspace{-10pt}
\end{figure}

Consider a primary communication system (e.g., multicarrier based)
over a wideband channel that is divided into $K$ non-overlapping
subchannels. At a particular time, some of the $K$ subchannels
might not be used by the primary users and are available for
opportunistic spectrum access. Multiuser orthogonal frequency
division multiplexing (OFDM) schemes are suitable candidates for
such a scenario since they make it convenient to nullify or
activate some portion of multiple narrow bands. We model the
detection problem over the subband $k$ as one to choose between
hypothesis $\mathcal{H}_{0,k}$ (``0''), which represents the
absence of primary signals, and hypothesis $\mathcal{H}_{1,k}$
(``1''), which represents the presence of primary signals. An
illustration where only some of the $K$ bands are occupied by
primary users is illustrated in Fig.
\ref{fig:sppool}. 
The crucial task of spectrum sensing is to sense the $K$ frequency
bands and identify spectral holes for opportunistic use. For
simplicity, we assume that the high-layer protocols guarantee that
all CRs keep quiet during the detection such that the main
spectral
power under detection is emitted by the primary users. 

Consider a multi-path fading environment, where $h(l)$, $l=0, 1,
\ldots, L-1$, denotes the discrete-time channel impulse response
between the primary transmitter and a CR receiver with $L$ equal
to the number of resolvable paths. The received baseband signal at
the CR front-end can be expressed as\vspace{-5pt}
\begin{equation}
r(n)= \sum_{l=0}^{L-1} h\left(l\right)  s\left(n-l\right)  + v(n),
\ \ n=0, 1, \ldots, N_0-1 \vspace{-8pt}
\end{equation}
where $s(n)$ represents the primary transmitted signal with cyclic
prefix at time $n$ and $v(n)$ is additive complex white Gaussian
noise with zero mean and variance $\sigma_v^2$, i.e., $v(n) \sim
\mathcal{CN}\left(0, \sigma_v^2 \right)$. In a multi-path fading
environment, the wideband channel exhibits frequency-selective
features and its discrete frequency responses are given by
\vspace{-5pt}
\begin{equation}
H_k = \frac{1}{\sqrt{N_0}} \sum_{n=0}^{L-1}  h(n) e^{-j 2 \pi
nk/N_0}, \ \ k=0, 1, \ldots, K-1 \vspace{-8pt}
\end{equation}
where $L \leq N_0$. We assume that the channel is slowly varying
and the channel frequency responses $\{H_k\}_{k=0}^{K-1}$ do not
vary much during a detection interval. In the frequency domain,
the received signal at each subchannel can be estimated by
computing its
discrete Fourier transform (DFT): 
\begin{equation}\vspace{-6pt}
R_k = \frac{1}{\sqrt{N_0}}\sum_{n=0}^{N_0-1} r(n)  e^{-j 2\pi n
k/N_0} = H_k S_k + V_k
\end{equation}
where $S_k$ is the primary signal at subchannel $k$ and
\begin{equation}
 V_k =
\frac{1}{\sqrt{N_0}} \sum_{n=0}^{L-1}  v(n) e^{-j 2 \pi nk/N_0}
\vspace{-8pt}
\end{equation}
is the received noise in the frequency domain. Note that $V_k \sim
\mathcal{CN}\left(0, \sigma_v^2 \right)$ since $v(n) \sim
\mathcal{CN}\left(0, \sigma_v^2 \right)$ and the DFT is a unitary
linear operation. Without loss of generality, we assume that the
transmitted signal $S_k$, channel gain $H_k$, and additive noise
$V_k$ are independent of each other.

To decide whether the $k$-th subchannel is occupied or not, we
test the following binary hypotheses:
\begin{align}
&\mathcal{H}_{0,k} :~ R_k = V_k, \ \ \ \ \ \ \ \ \ \ \ \ \ \ \ \
k=0, 1, \ldots, K-1\nonumber
\\
&\mathcal{H}_{1,k}:~ R_k = H_k S_k +V_k,   \ \ \ \ \ \ k=0, 1,
\ldots, K-1 \vspace{-9pt}
\end{align}
For each subchannel $k$, we compute the test statistic as the sum
of received energy over an interval of $M$ samples,
i.e.,\vspace{-3pt}
\begin{equation}\vspace{-5pt}
Y_k = \sum_{m=0}^{M-1} \left|R_k(m) \right|^2  \ \ \ \ \ \ k=0, 1,
\ldots, K-1
\end{equation}
and the decision rule is given by \vspace{-3pt}
\begin{equation} \label{eqn:NarrowBand_Detection}
Y_k  \begin{array}{c} \mathcal{H}_{1,k} \\ \gtreqless \\
\mathcal{H}_{0,k} \end{array}  \gamma_k \vspace{-6pt}
\end{equation}
where $\gamma_k$ is the corresponding decision threshold. For
simplicity, we assume that the transmitted signal at each
subchannel has unit power, i.e.,  $\mathbb{E} \left(|S_k|^2\right)
=1$; this assumption holds when the primary radios adopt uniform
power transmission strategies given no channel knowledge at the
transmitter side. 

According to the central limit theorem for large $M$, $Y_k$ is
asymptotically normally distributed with mean
\begin{equation} \label{eqn:uc_bar}
\mathbb{E} \left( Y_{k} \right) = \left\{
\begin{array}{ll}
    M \sigma_v^2    & \ \ \ \mathcal{H}_{0,k} \vspace{2pt}\\
    M \left(\sigma_v^2 + |H_k|^2 \right)  & \ \ \  \mathcal{H}_{1,k} \\
\end{array}%
\right. \vspace{-6pt}
\end{equation}
and variance
\begin{align}\vspace{-6pt}
\mathrm{Var} \left(Y_k\right)  = \left\{
\begin{array}{ll}
    2M \sigma_v^4    & \ \ \ \mathcal{H}_{0,k} \vspace{2pt} \\
    2M \left(\sigma_v^2 + 2|H_k|^2 \right)\sigma_v^2   & \ \ \  \mathcal{H}_{1,k} \\
\end{array}%
\right.
\end{align}
for $k=0, 1, \ldots, K-1$. Thus, assuming large $M$, we have $Y_k
\sim \mathcal{N} \left( \mathbb{E} \left(Y_{k}\right),
\mathrm{Var} \left(Y_k\right) \right)$.

Using the decision rule in (\ref{eqn:NarrowBand_Detection}), the
probabilities of false alarm and detection at subchannel $k$ can
be respectively calculated approximately as
\begin{align} \label{eqn:P_f} \vspace{-8pt}
P_f^{(k)}(\gamma_k) = \mathrm{Pr} \left(Y_k > \gamma_k |
\mathcal{H}_0 \right) = Q \left( \frac{\gamma_k-M \sigma_v^2
}{\sigma_v^2 \sqrt{2M}}\right) \vspace{-6pt}
\end{align}
and
\begin{equation} \label{eqn:p_d} \vspace{-6pt}
P_d^{(k)}(\gamma_k) = Q \left(\frac{\gamma_k-M \left(\sigma_v^2 +
|H_k|^2 \right) }{\sigma_v \sqrt{2M \left(\sigma_v^2 + 2|H_k|^2
\right)}}\right)
\end{equation}
where $Q$ denotes the tail probability of the standard normal
distribution. The choice of threshold $\gamma_k$ leads to a
tradeoff between the probabilities of false alarm and miss
$P_m=1-P_d$. Specifically, a higher threshold will result in a
smaller probability of false alarm but a larger probability of
miss, and vice versa.

%
%
\vspace{-10pt}
\section{Spatial-Spectral Joint Detection}\label{sec:dist_coop}
\label{sec:framework} \vspace{-8pt}

Suppose that $N$ spatially distributed cognitive radios
collaboratively sense a wide frequency band. By combining the
local statistics from individual cognitive radios at the fusion
center, which can be one of the CRs, the network can make a better
decision on the presence or absence of primary signals on each of
the $K$ subchannels. The cooperation assumes a separate control
channel, through which the statistics of individual CRs are
transmitted to the fusion center. Let $Y_k(n)$ denote the received
energy in the $k$-th subchannel at cognitive radio $n$. For each
subchannel, these statistics can be written in a vector as
$\boldsymbol{Y}_k=\left[Y_{k}(0), Y_{k}(1), \ldots, Y_{k}(N-1)
\right]^T$.


To exploit the spatial diversity, we linearly combine the summary
statistics from spatially distributed cognitive radios at each
subchannel $k$ to obtain a final test statistic: \vspace{-6pt}
\begin{equation}
z_k = \sum_{n=0}^{N-1} w_k(n) Y_k(n) = \boldsymbol{w}_k^T
\boldsymbol{Y}_k \vspace{-6pt}
\end{equation}
where $\boldsymbol{w}_k=\left[w_k(0), w_k(1), \ldots, w_k(N-1)
\right]^T$ are the combining coefficients for subchannel $k$,
which can be compactly written in a matrix as
$\boldsymbol{W} = \left[\boldsymbol{w}_0 \ \boldsymbol{w}_1 \ \ldots \ \boldsymbol{w}_{K-1} \\
\right] $. Note that $w_k(n)\geq 0$, for every $k$ and $n$.

Since the entries in $\boldsymbol{Y}_k$ are normally distributed,
the test statistics $\{z_k\}_{k=0}^{K-1}$ are also normally
distributed with means \vspace{-6pt}
\begin{equation}
\mathbb{E} \left( z_k \right)= \left\{
\begin{array}{ll}
    M \sigma_v^2 \boldsymbol{w}_k^T \boldsymbol{1}   & \ \ \ \mathcal{H}_{0,k} \vspace{4pt}\\
    M \boldsymbol{w}_k^T \left(\sigma_v^2  \boldsymbol{1}  + \boldsymbol{G}_k \right)  & \ \ \  \mathcal{H}_{1,k} \\
\end{array}%
\right. \vspace{-6pt}
\end{equation}
where $\boldsymbol{1}$ is an all-one vector, and
variances\vspace{-6pt}
\begin{equation}
\mathrm{Var} \left(z_k\right) = \left\{
\begin{array}{ll}
    2M \sigma_v^4 \boldsymbol{w}_k^T \boldsymbol{w}_k   & \ \ \ \mathcal{H}_{0,k} \vspace{4pt}\\
    2M \sigma_v^2  \boldsymbol{w}_k^T \left[\sigma_v^2  \boldsymbol{I}  + 2\mathrm{diag}(\boldsymbol{G}_k) \right]\boldsymbol{w}_k  & \ \ \  \mathcal{H}_{1,k} \\
\end{array}%
\right.
\end{equation}
where $\boldsymbol{G}_k=\left[|H_k(0)|^2, |H_k(1)|^2, \ldots,
|H_k(N-1)|^2 \right]^T$ are the squared magnitudes of the channel
gains between the primary transmitter and the $N$ CR receivers for
subchannel $k$.

In order to decide the presence or absence of the primary signal
in subchannel $k$, we use the following binary test \vspace{-6pt}
\begin{equation} \label{eqn:distDetection}
z_k  \begin{array}{c} \mathcal{H}_{1,k} \\ \gtreqless \\
\mathcal{H}_{0,k} \end{array}  \gamma_k, \ \ \ \ k=0, 1, \ldots,
K-1. \vspace{-6pt}
\end{equation}
Accordingly, the detection performance in terms of the
probabilities of false alarm and detection are given by
\vspace{-4pt}
\begin{equation}\label{eqn:P_f_W_gamma}
P_f^{(k)}(\boldsymbol{w}_k, \gamma_k) = Q\left( \frac{\gamma_k-M
\sigma_v^2 \boldsymbol{w}_k^T \boldsymbol{1}  }{\sigma_v^2
 \sqrt{2M \boldsymbol{w}_k^T \boldsymbol{w}_k }}\right) \vspace{-10pt}
\end{equation}
and \vspace{-8pt}
\begin{equation}\label{eqn:P_d_W_gamma}
P_d^{(k)}(\boldsymbol{w}_k, \gamma_k) = Q\left( \frac{\gamma_k- M
\boldsymbol{w}_k^T \left(\sigma_v^2  \boldsymbol{1}  +
\boldsymbol{G}_k \right) }{\sigma_v \sqrt{2M  \boldsymbol{w}_k^T
\left[\sigma_v^2  \boldsymbol{I}  + 2
\mathrm{diag}(\boldsymbol{G}_k) \right]\boldsymbol{w}_k }}\right)
\end{equation}
For compactness of notation, we collect the probabilities of false
alarm and detection over the $K$ subchannels into vectors
$\boldsymbol{P}_f (\boldsymbol{W}, \boldsymbol{\gamma})$ and
$\boldsymbol{P}_d (\boldsymbol{W}, \boldsymbol{\gamma})$. Thus,
the probabilities of miss can be represented as $\boldsymbol{P}_m
(\boldsymbol{W}, \boldsymbol{\gamma}) =
\boldsymbol{1}-\boldsymbol{P}_d (\boldsymbol{W},
\boldsymbol{\gamma})$.


Our goal is to maximize the opportunistic rate while meeting some
constraints on the interference to the primary communication
system. Let $r_k$ denote the throughput achievable over the $k$-th
subchannel if used by cognitive radios, and
$\boldsymbol{r}=\left[r_0, r_1, \ldots, r_{K-1}\right]^T$. Since
$1-P_f^{(k)}$ measures the opportunistic spectrum utilization of
subchannel $k$, we define the aggregate opportunistic throughput
capacity as \vspace{-6pt}
\begin{equation}
R\left(\boldsymbol{W}, \boldsymbol{\gamma}\right) =
\boldsymbol{r}^T
\left[\boldsymbol{1}-\boldsymbol{P}_f(\boldsymbol{W},
\boldsymbol{\gamma}) \right]. \vspace{-6pt}
\end{equation}

For a widband primary system, the impact of interference induced
by cognitive devices can be characterized by a relative priority
vector over the $K$ subchannels, i.e.,
$\boldsymbol{c}=\left[c_{0}, c_{1}, \ldots, c_{K-1}\right]^T$,
where $c_k$ indicates the cost incurred if the primary user at
subchannel $k$ is interfered with. As such, we define the
aggregate interference to the primary user as $ \boldsymbol{c}^T
\boldsymbol{P}_m (\boldsymbol{W}, \boldsymbol{\gamma}) $.
Consequently, the spatial-spectral joint detection problem is
formulated as
\begin{align}\vspace{-6pt}
&\max_{\boldsymbol{W}, \boldsymbol{\gamma}}   \ \ \ \ \ R\left(\boldsymbol{W}, \boldsymbol{\gamma}\right) \hfill & \hfill{(\mathrm{P}1)} \nonumber \\
&\ \ \    \mathrm{s.t.} \ \ \ \ \   \boldsymbol{c}^T
\boldsymbol{P}_m (\boldsymbol{W}, \boldsymbol{\gamma}) \leq
\varepsilon
\label{eqn:const_total_interference}  \\
& \ \ \ \ \ \  \ \ \ \ \ \  \boldsymbol{P}_m (\boldsymbol{W}, \boldsymbol{\gamma}) \preceq \boldsymbol{\alpha} \label{eqn:Pm_const_W_gamma}  \\
& \ \ \ \ \ \  \ \ \ \ \ \  \boldsymbol{P}_f (\boldsymbol{W},
\boldsymbol{\gamma}) \preceq \boldsymbol{\beta}
\label{eqn:Pf_const_W_gamma}\vspace{-6pt}
\end{align}
where $ \boldsymbol{\alpha}=[\alpha_0, \ldots, \alpha_{K-1}]^T$
and $\boldsymbol{\beta}=[\beta_0, \ldots, \beta_{K-1}]^T$.

Finding the exact solution for the above problem is difficult
since for any $k$, $P_f^{(k)}(\boldsymbol{w}_k, \gamma_k)$ and
$P_d^{(k)}(\boldsymbol{w}_k, \gamma_k)$ are neither convex nor
concave functions according to (\ref{eqn:P_f_W_gamma}) and
(\ref{eqn:P_d_W_gamma}). To jointly optimize $\boldsymbol{W}$ and
$\boldsymbol{\gamma}$, we can show that ($\mathrm{P}1$) can be
reformulated into an equivalent form with convex constraints and
an objective function lower bounded by a concave function under
the following conditions:
\begin{equation}\label{eqn:pratical_conditions}
0 < \alpha_k \leq \frac{1}{2} \ \ \mathrm{and} \ \ 0< \beta_k \leq
\frac{1}{2}, \ \ k=0, \ldots, K-1.
\end{equation}
Through maximizing the lower bound of the objective function, we
are able to obtain a good approximation to the optimal solution of
the original problem.

First, we show how to transform the nonconvex constraints in
(\ref{eqn:Pm_const_W_gamma}) and (\ref{eqn:Pf_const_W_gamma}) into
convex constraints by exploiting the monotonicity of the
$Q$-function. Substituting (\ref{eqn:P_f_W_gamma}) into the
constraint (\ref{eqn:Pf_const_W_gamma}), we have
\begin{equation}\label{eqn:const_convex_beta}
Q^{-1}(\beta_k) \sqrt{2M \boldsymbol{w}_k^T \boldsymbol{w}_k} \leq
\frac{\gamma_k}{\sigma_v^2 }-M \boldsymbol{w}_k^T \boldsymbol{1}
\vspace{-2pt}
\end{equation}
where $Q^{-1}(\beta_k) \geq 0$ given $\beta_k \leq 1/2$. From
(\ref{eqn:P_d_W_gamma}), the constraint
(\ref{eqn:Pm_const_W_gamma}) can be expressed as
\begin{equation}\label{eqn:const_convex_alpha}
 \sqrt{2M  \boldsymbol{w}_k^T
\left[\sigma_v^2  \boldsymbol{I}  + 2
\mathrm{diag}(\boldsymbol{G}_k) \right]\boldsymbol{w}_k }  \leq
\frac{\gamma_k -M \boldsymbol{w}_k^T \left(\sigma_v^2
\boldsymbol{1}  + \boldsymbol{G}_k \right)}{\sigma_v
Q^{-1}(1-\alpha_k)} \vspace{-1pt}
\end{equation}
given $\alpha_k \leq 1/2 $ and $Q^{-1}(1-\alpha_k) \leq 0$. Since
the left-hand side on the constraint (\ref{eqn:const_convex_beta})
is convex and the right hand side is linear in ($\gamma_k$,
$\boldsymbol{w}_k$), (\ref{eqn:const_convex_beta}) defines a
convex set for $(\gamma_k, \boldsymbol{w}_k )$. Similarly,
(\ref{eqn:const_convex_alpha}) is also a convex constraint.


Then, we reformulate ($\mathrm{P}1$) by introducing a new variable
\begin{equation}\mu_k =  \sigma_v \sqrt{2M \boldsymbol{w}_k^T
\left[\sigma_v^2 \boldsymbol{I}  +
2\mathrm{diag}(\boldsymbol{G}_k) \right]\boldsymbol{w}_k }.
\end{equation}
By defining $\gamma_k' = \gamma_k /\mu_k$ and $\boldsymbol{w}_k' =
\boldsymbol{w}_k/\mu_k $, the constraints
(\ref{eqn:const_convex_beta}) and (\ref{eqn:const_convex_alpha})
can be further written as
\begin{equation}\label{eqn:const_convex_beta_prime}
\ Q^{-1}(\beta_k) \sqrt{2M {\boldsymbol{w}_k'}^T
\boldsymbol{w}_k'} \leq \frac{\gamma_k'}{\sigma_v^2 }- M
\boldsymbol{1}^T \boldsymbol{w}_k'
\end{equation} \vspace{-8pt}
and
\begin{equation}\label{eqn:const_convex_alpha_prime}
\gamma_k' -M
 \left(\sigma_v^2 \boldsymbol{1}  +
\boldsymbol{G}_k \right)^T \boldsymbol{w}_k' \leq \sigma_v
Q^{-1}(1-\alpha_k). \vspace{-6pt}
\end{equation}
Note that (\ref{eqn:const_convex_alpha_prime}) is actually a
linear constraint in ($\gamma_k'$, $\boldsymbol{w}_k'$). The
constraint (\ref{eqn:const_total_interference}) now becomes
\begin{equation}
\boldsymbol{1}^T \boldsymbol{c} - \sum_{k=0}^{K-1} c_k Q
\left(\gamma_k'- M
 \left(\sigma_v^2 \boldsymbol{1}  +
\boldsymbol{G}_k \right)^T \boldsymbol{w}_k'\right) \leq
\varepsilon \label{eqn:const_const1},
\end{equation}
which can be shown to be convex by the following result.

\begin{lemma}\label{lemma:3}
If $\gamma_k' \leq M
 \left(\sigma_v^2 \boldsymbol{1}  +
\boldsymbol{G}_k \right)^T \boldsymbol{w}_k'$, then the function
$Q \left(\gamma_k'- M
 \left(\sigma_v^2 \boldsymbol{1}  +
\boldsymbol{G}_k \right)^T \boldsymbol{w}_k'\right)$ is concave in
$\{\gamma_k', \boldsymbol{w}_k' \}$.
\end{lemma} \vspace{-8pt}

By changing the variables $\boldsymbol{W}'$ and
$\boldsymbol{\gamma}'$, $P_f^{(k)}(\boldsymbol{w}_k, \gamma_k) $
can be expressed as
\begin{equation}
Q\left[ \left(\frac{\gamma_k'}{\sigma_v^2}- M \boldsymbol{1}^T
\boldsymbol{w}_k' \right) \sqrt{ \sigma_v^2 + \frac{2
{\boldsymbol{w}_k'}^T \mathrm{diag}(\boldsymbol{G}_k)
\boldsymbol{w}_k' }{{\boldsymbol{w}_k'}^T \boldsymbol{w}_k'}}
\right]
\end{equation}
From the the Rayleigh-Ritz theorem \cite{Sayed2003}, we have
\begin{equation}
\min_{ n } \left| H_k(n) \right|^2 \leq \frac{
{\boldsymbol{w}_k'}^T \mathrm{diag}(\boldsymbol{G}_k)
\boldsymbol{w}_k' }{{\boldsymbol{w}_k'}^T \boldsymbol{w}_k'} \leq
\max_{n }  \left| H_k(n) \right|^2
\end{equation}
Define a new function\vspace{-4pt}
\begin{align}
g_k \left(\gamma_k', \boldsymbol{w}_k' \right) &\stackrel{\Delta}
{=} Q \left[ \left(\frac{\gamma_k'}{\sigma_v^2}- M
\boldsymbol{1}^T\boldsymbol{w}_k' \right) \sqrt{\sigma_v^2
 +  2 \min_{ n } |H_k(n)|^2 } \right]
\end{align}\vspace{-4pt}
which can be shown to be convex by the following result.
\begin{lemma}
If $\gamma_k' \geq \sigma_v^2 M  \boldsymbol{1}^T
\boldsymbol{w}_k'$, then the function $g_k(\gamma_k',
\boldsymbol{w}_k' )$ is convex in $\{\gamma_k', \boldsymbol{w}_k'
\}$.
\end{lemma}\vspace{-3pt}

Since $P_f^{(k)}(\boldsymbol{w}_k, \gamma_k) \leq g_k(\gamma_k',
\boldsymbol{w}_k' )$, the objective function in ($\mathrm{P}1$)
can be lower bounded by $ \sum_{k=0}^{N-1} r_k\left[1- g_k
\left(\gamma_k', \boldsymbol{w}_k' \right) \right]$, which is a
concave function. Thus, an efficient suboptimal method to solve
($\mathrm{P}1$) is to maximize the lower bound of its objective
function, i.e.,
\begin{align} & \max_{\boldsymbol{W}', \boldsymbol{\gamma}'}   \ \ \ \ \ \sum_{k=0}^{N-1} r_k\left[1- g_k \left(\gamma_k',
\boldsymbol{w}_k' \right) \right]
\ \ \ \ \ \ \ \ \ \ \ \ \ \ \ \ \ \ \ \hfill{(\mathrm{P}2)} \nonumber \\
& \ \ \   \mathrm{st.} \ \ \ \ -\sum_{k=0}^{K-1} c_k Q
\left[\gamma_k'- M
 \left(\sigma_v^2 \boldsymbol{1}  +
\boldsymbol{G}_k \right)^T \boldsymbol{w}_k'\right] \leq
\varepsilon- \boldsymbol{1}^T\boldsymbol{c}
\nonumber \\
& \ \ \ \ \ \  \ \ \ \ \ \  Q^{-1}(\beta_k) \sqrt{2M
{\boldsymbol{w}_k'}^T \boldsymbol{w}_k'} \leq
\frac{\gamma_k'}{\sigma_v^2 }- M \boldsymbol{1}^T
\boldsymbol{w}_k' \ \nonumber \\
& \ \ \ \ \ \  \ \ \ \ \ \ \gamma_k' -M
 \left(\sigma_v^2 \boldsymbol{1}  +
\boldsymbol{G}_k \right)^T \boldsymbol{w}_k' \leq \sigma_v
Q^{-1}(1-\alpha_k)  \nonumber.
\end{align}
Implied by the practical conditions in
(\ref{eqn:pratical_conditions}), this problem is a convex
optimization problem and can be solved efficiently.

\vspace{-10pt}
\section{SIMULATIONS}
\label{sec:sim}\vspace{-6pt}

\begin{figure}
\centering
\includegraphics[width=3in]{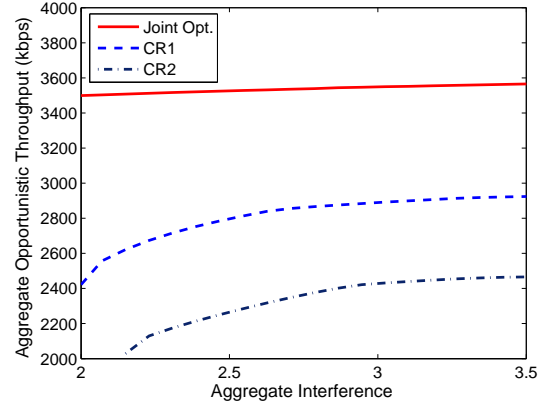}
\caption{\small Aggregate opportunistic throughput capacity vs.
the constraint on the aggregate induced interference.}
\label{fig:Coop_Th_vs_Int} \vspace{-10pt}
\end{figure}

Suppose that two CRs cooperatively sense a multiband OFDM system
with $8$ subbands. For each subband, it is expected that the
opportunistic spectrum utilization is at least $50\%$, i.e.,
$\beta_k = 0.5$, and the probability that the primary user is
interfered is at most $\alpha_k=0.1$. It is assumed that
$\sigma_v^2=1$ and $M=100$. Other parameters are given in Table
\ref{table:ex2}. Fig. \ref{fig:Coop_Th_vs_Int} shows result of
solving ($\mathrm{P}2$), which maximizes the opportunistic
throughput subject to the constraints on the interference. We
observe that the joint detection results in much higher
opportunistic throughput than the algorithms without cooperation.
Note that the increase in the throughput of the joint optimization
scheme becomes rather slow as we relax the interference constraint
because the interaction between $\boldsymbol{\gamma}$ and
$\boldsymbol{W}$ pushes the system to an operating point at which
the throughput is more limited by $\boldsymbol{\beta}$ than by
$\varepsilon$.

\begin{table}
\renewcommand{\arraystretch}{1.2}
\caption{Parameters used in simulations}
\vspace{4pt}\label{table:ex2} \centering
\begin{tabular}{c||c|c|c|c|c|c|c|c}
\hline
\bfseries \small  $\boldsymbol{G}(0)$& \small .17 & \small .21 & \small .27 & \small .14 & \small .37 & \small .38 & \small .49 & \small .33 \\
\hline
\bfseries \small $\boldsymbol{G}(1)$& \small .21 & \small .17 & \small .21 & \small .21 & \small .17 & \small .43 & \small .15 & \small .35 \\
\hline
\small $\boldsymbol{r}$ & \small 356 & \small 327 & \small 972 & \small 806 & \small 755 & \small 68 & \small 720 & \small 15  \\
\hline
\small $\boldsymbol{c}$ & \small .71 & \small 5.95 & \small 3.91  & \small 4.21 & \small .44 & \small 2.03 & \small .58 & \small 2.85  \\
\hline
\end{tabular} \\
\end{table}

\vspace{-8pt}
\section{CONCLUSION}
\label{sec:cls}\vspace{-8pt}

In this paper, we have proposed a spatial-spectral joint detection
framework for distributed wideband spectrum sensing in cognitive
radio networks, within which the cooperation among spatially
distributed cognitive radios is optimized over multiple frequency
bands. By exploiting the inherent structure of the formulation, we
have developed suboptimal but efficient solutions for the
non-convex optimization problem. This paper establishes important
principles for the design of distributed wideband spectrum sensing
algorithms in cognitive radio networks.



\vspace{-9pt}

\small

\bibliographystyle{IEEEbib}
\bibliography{strings,refs}

\end{document}